\documentstyle[aps, amsfonts, amsmath, epsfig ,rotate]{revtex}
\newcommand{\bef}{\begin{figure}}
\newcommand{\eef}{\end{figure}}

\newcommand{\leb}{\left(}
\newcommand{\rib}{\right)}
\newcommand{\bei}{\begin{itemize}}
\newcommand{\eei}{\end{itemize}}
\newcommand{\bea}{\begin{eqnarray}}

\newcommand{\smsum}{\mbox{$\sum$}}
\newcommand{\sig}{\sigma}

\newcommand{\eea}{\end{eqnarray}}
\newcommand{\bequ}{\begin{equation}}
\newcommand{\eequ}{\end{equation}}

\begin{document}

\title{Non--equilibrium stationary state of a two-temperature spin chain}
\author{F.~Schm\"user\footnote{Present address: MPI f\"ur Physik komplexer
Systeme, N\"othnitzer Str.~38, 01187 Dresden, Germany. e-mail: frank@mpipks-dresden.mpg.de}
and B.~Schmittmann\footnote{e-mail: schmittm@vt.edu}}
\address{Center for Stochastic Processes in Science and Engineering, and 
Department of Physics, Virginia Tech, Blacksburg, Virginia 24061-0435, USA}
\maketitle
\begin{abstract}
A kinetic one--dimensional Ising model is coupled to two heat baths, such
that spins at even (odd) lattice sites experience a temperature $T_{e}$ ($%
T_{o}$). Spin flips occur with Glauber-type rates generalised to the case of
two temperatures. Driven by the temperature differential, the spin chain
settles into a non-equilibrium steady state which corresponds to the
stationary solution of a master equation. We construct a perturbation
expansion of this master equation in terms of the temperature difference
and compute explicitly the first two corrections to the equilibrium
Boltzmann distribution. The key result is the emergence of additional spin
operators in the steady state, increasing in spatial range and order of spin
products. We comment on the violation of detailed balance and entropy
production in the steady state.
\end{abstract}

\pacs{05.70.Ln, 05.50.+q, 02.50.-r}
\section{Introduction}
In recent years, considerable interest has focused on non-equilibrium
stationary states (NESS)\ in interacting many-body systems, induced by
deterministic or stochastic microscopic dynamics \cite{ruelle,schmizi}. In
the stochastic case, the starting point is typically a master equation,
i.e., a continuity equation for the (time-dependent) configurational
probabilities. Different models are characterised by their 
microscopic transition rates. Both,
equilibrium and non-equilibrium stationary states, have
time-independent macroscopic observables. However, only in the equilibrium
case it is possible to compute these observables without explicit reference
to the imposed dynamics, in the framework of Gibbs ensembles. In stark
contrast, NESS and their properties depend generically on the details of the
dynamic rules. Since a unifying theoretical description of NESS\ is still
lacking, most progress to date is made by studying specific models.

In some of the simplest models, a NESS\ is established by driving the system
with an external field. A characteristic example is the (fully periodic)
Ising lattice gas, subject to a uniform electric field which induces a
nonzero particle current \cite{KLS}. A second class of models involves two
temperature baths. These can be coupled either to the system boundaries \cite%
{T-grad,luc}, or act throughout the bulk: in the latter case, each temperature
bath controls a (translationally invariant) subset of configurational
transitions. For instance, in an Ising lattice gas, particle-hole exchanges
along certain spatial axes might be coupled to a higher temperature \cite%
{KETT,KettRG}. Or, in the non-conserved case, spin-flips on a selected
sub-lattice might occur at a different temperature \cite{SFTT,blote}. In all
of these cases, equilibrium can be recovered upon letting a key parameter
vanish - e.g., the strength of the bias, or the temperature difference. In
this spirit, these models allow us to probe the effect of non-equilibrium
perturbations on equilibrium properties, and we may ask to what extent \emph{%
universal} (i.e., long-wavelength, long-time) features are affected. One
finds, generically, that non-equilibrium perturbations are far more
relevant, in a renormalisation group sense, in systems with conservation
laws \cite{KettRG,KLSRG,UCT}. In contrast, it can be shown that Ising-like
models with \emph{non-conserved} dynamics remain in the universality class
of the equilibrium Ising model, even if the usual $\mathbb{Z}_{2}$-symmetry
of the Ising model is broken \cite{cr-beh-noncons}. While these results are
clearly of major theoretical interest, they do not reveal how NESS
configurational probabilities differ from their equilibrium counterparts:
How, for instance, is the Boltzmann distribution for the equilibrium Ising
model modified, when a second, different temperature bath is coupled to some
spin-flips or -exchanges? Clearly, such modifications must be severe if universal
behaviour is to be affected. Yet, even if long-wavelength, long-time
properties remain effectively Ising-like, fundamental changes in the
configurational probabilities should arise from the breaking of detailed
balance.

It is with these motivations, questions and expectations in mind, that we
turn to the arguably simplest non-equilibrium Ising-type model, namely, a
one-dimensional \emph{interacting} spin chain with spin-flip dynamics. The
rates are a simple but nontrivial generalisation of the well-studied Glauber %
\cite{glau} rates: spins at odd (even) sites are coupled to a temperature $%
T_{o}$ ($T_{e}$). In dimensions $d\geq 2$, this model \cite{SFTT,blote}
exhibits an order-disorder phase transition which should be in the Ising
universality class, according to renormalisation group arguments \cite%
{cr-beh-noncons}. Detailed Monte Carlo simulations in $d=2$ \cite{blote}
confirm this expectation. In $d=1$, an exact solution for the two spin
correlation functions has been found \cite{raczzia}. In the
following, we seek an analytic expression for the steady state probability
distribution, i.e., the solution of the master equation, for this model.
While we have not been able to find an exact solution, our perturbative
analysis provides some insight into how the equilibrium distribution is
modified when a (small) temperature difference is present.

As energy is fed into a non--equilibrium stationary state from one
reservoir, it must also be dissipated: in general, a NESS constantly
produces ``entropy''. Possible definitions of ``entropy'' and ``entropy
production'' in non--equilibrium systems have been discussed intensely in
recent years, from the viewpoints of both stochastic and deterministic
dynamics \cite{ruellep,maes1}. However, different perspectives have been
developed rather independently, so that the relations between them still
remain somewhat unclear. Therefore, we find it worthwhile to compare results
from different approaches to entropy production, using our model.

We conclude with two comments: First, while some analytic results for
steady-state distributions are available, they are confined to two classes
of systems: first, $d=1$ lattice gas models, restricted to excluded volume
interactions, such as the asymmetric exclusion process and its relatives %
\cite{schuetz,priv,LS}, and second, very special ($d=1$) spin systems whose
master equations are solved by the Ising Boltzmann factor \cite{KLS,matt}.
Second, even though our main emphasis rests on fundamental aspects, we note
briefly that two-temperature systems of this type can be established in real
systems, and maintained for significant amounts of time. For example, via
nuclear magnetic resonance in an external magnetic field, a lattice of
nuclei in a solid can be prepared to have a certain \emph{spin temperature} %
\cite{abra}. Thus, in a crystal with two sub-lattices of different nuclei
the temperatures of these sub-lattices can differ.

This article is organised as follows. We begin by introducing our model and
its master equation. The following section describes our perturbative
approach towards finding the stationary distribution. We then compute the
two leading orders, aiming to highlight some
generic properties of the perturbation series. We next turn to the
discussion of two more general questions, namely, first, how spin flips
violate the detailed balance condition, and second, whether different
definitions for the entropy production rate lead to the same results, for
our model. We conclude with an outlook and some open questions.

\section{The model}

\label{secmodel} Our model is defined on a one-dimensional spin chain, of
length $N$ (with $N$ even). Each site $i$ carries a spin variable, $\sigma
_{i}$, which can take the values $\pm 1$. We impose periodic boundary
conditions, so that $\sigma _{N+1}=\sigma _{1}$. Spins at even (odd) lattice
sites are coupled to a heat bath at temperature $T_{e}$ ($T_{o}$). Thus, a
configuration $\{\sigma \}=\{\sigma _{1},\,\sigma _{2},\dots ,\sigma _{N}\}$
evolves into a new configuration by flipping a randomly selected spin, e.g.,
spin $\sigma _{i}$, with a rate \cite{raczzia} 
\begin{equation}
w_{i}(\sigma _{i}\rightarrow -\sigma _{i})=1-\frac{\gamma _{i}}{2}\sigma
_{i}(\sigma _{i-1}+\sigma _{i+1})\;,  \label{eqrates}
\end{equation}
where 
\begin{equation}
\quad \gamma _{i}=\left\{ 
\begin{array}{l}
\;\gamma _{e}=\tanh (2J/k_{B}T_{e})\,,\quad \quad i\;\;{even} \\ 
\;\gamma _{o}=\tanh (2J/k_{B}T_{o})\,,\quad \quad i\;\;{odd}\quad .%
\end{array}
\right.  \label{gamma_i}
\end{equation}
Here, $J$ denotes a nearest-neighbour exchange coupling between spins. From
now on, we use dimensionless units for inverse temperature, i.e., $\beta
_{e}\equiv J / (k_{B}\, T_{e})$, etc. 
The rates respect the usual $\mathbb{Z}_{2}$%
-symmetry of the Ising model. The full stochastic dynamics of the system can
be cast in terms of a master equation for the time-dependent
configurational probability $p(\{\sigma \};t)$ 
\begin{equation}
\partial _{t}p(\{\sigma \};t)=\sum_{i=1}^{N}\left[ -w_{i}(\sigma
_{i}\rightarrow -\sigma _{i})p\left( \{\sigma \};t\right)
+w_{i}(-\sigma _{i}\rightarrow \sigma _{i})p\left( \{\sigma ^{\lbrack
i]}\};t\right) \right]  \label{me}
\end{equation}
Here, $\{\sigma ^{\lbrack i]}\}$ differs from $\{\sigma \}$ by a flip of the 
$i$-th spin. A trivial time scale has been set to unity.

An alternate description of our spin chain is the usual defect (domain wall)
picture. A unique defect configuration $\{\alpha \}$ is associated with a
pair of spin configurations, namely $\{\sigma \}$ and its image under $%
\mathbb{Z}_{2}$, $\{-\sigma \}$. The defects are located on the \emph{dual}
lattice (i.e., the bonds), with $\alpha _{i}=0$ ($1$) if $\sigma _{i}\sigma
_{i+1}=1$ ($-1$). A spin flip $\sigma _{i}\rightarrow -\sigma _{i}$ in the
spin chain $\{\sigma \}$ can give rise to three different types of processes
in the defect state $\{\alpha \}$ depending on the three neighbouring spins $%
\{\sigma _{i-1},\,\sigma _{i},\,\sigma _{i+1}\}$: (a) if $\sigma
_{i-1}=-\sigma _{i}=\sigma _{i+1}$, two adjacent defects annihilate each
other, with rate $1+\gamma _{e}$ if $i$ is even ($1+\gamma _{o}$ if $i$ is
odd); (b) if $\sigma _{i-1}=\sigma _{i}=\sigma _{i+1}$, two adjacent defects
are generated simultaneously, with rate $1-\gamma _{e}$ if $i$ is even ($%
1-\gamma _{o}$ if $i$ is odd); and (c) if $\sigma _{i-1}=-\sigma _{i+1}$, a
defect diffuses to a neighbouring lattice site, with rate $1$.~

Clearly, for uniform temperature $T\equiv T_{e}=T_{o}$, this model reduces
to the well-known Glauber dynamics \cite{glau}, with homogeneous rates $\bar{%
w}(\sigma _{i}\rightarrow -\sigma _{i})$. The associated steady-state
solution is just the Boltzmann distribution, 
\begin{equation}
\lim_{t\rightarrow \infty }p(\{\sigma \};t)=\frac{1}{Z}\;\exp \left(
-H/k_{B}T\right) \equiv \frac{1}{Z}q_{o}\left( \{\sigma \}\right) \;
\label{ess}
\end{equation}
where $H$ is the Ising Hamiltonian for a spin chain, 
\begin{equation}
H\left( \{\sigma \}\right) =-J\sum_{i=1}^{N}\,\sigma _{i}\sigma _{i+1}\quad
\label{isingham}
\end{equation}
and $Z$ is the (canonical) partition function. Clearly, $H\left( \{\sigma
\}\right) $ just counts the number of defects, 
$n(\{\sigma \})\equiv\sum_{i=1}^{N}\alpha _{i}$, 
in configuration $\{\sigma \}$ via $J^{-1}H(\{\sigma \})=2n(\{\sigma \})-N$.
The equilibrium distribution $q_{0}\left( \{\sigma \}\right) /Z$ will serve
as the unperturbed reference solution for our perturbative analysis, to
be outlined below. For later reference, we note that detailed balance holds
in the equilibrium case, i.e., 
\begin{equation}
\bar{w}(\sigma _{i}\rightarrow -\sigma _{i})\;q_{0}\left( \{\sigma \}\right)
=\bar{w}(-\sigma _{i}\rightarrow \sigma _{i})\;q_{0}\left( \{\sigma
^{[i]}\}\right) \quad  \label{eqdetbal}
\end{equation}
for any spin-flip.

Returning to the case of two temperatures, it is easy to see that the (time
continuous) Markov process, characterised by Eqn.~(\ref{me}) is ergodic,
since every configuration $\{\sigma \}$ can be reached after a finite time
from every other configuration $\{\sigma ^{\prime }\}$. Hence, any initial
condition will converge, for $t\to \infty $, to the unique stationary
solution, $q\left( \{\sigma \}\right) $. The determination of $q\left(
\{\sigma \}\right) $ is our goal here. Being the full configurational
probability distribution, it contains \emph{all} information about the
system in its non-equilibrium steady state.

Since our NESS violates detailed balance (cf.~Section 4), the determination
of $q\left( \{\sigma \}\right) $ is highly non--trivial. As a starting
point, we restrict ourselves to the regime where the two temperatures differ
only slightly from each other, so that perturbative methods can be invoked.
In particular, we are interested which new spin operators are induced by the
coupling to two heat baths. Writing the stationary distribution as 
\begin{equation}
q\left( \{\sigma \}\right) \equiv \frac{1}{\tilde{Z}}\;\exp \left( V\left( \{\sigma
\}\right) \right) \;,  \label{eqpotenne}
\end{equation}
we seek the ``potential'' function $V\left( \{\sigma \}\right) $ in
perturbation theory ($\tilde{Z}$ is determined through the normalisation condition).

\section{Perturbation theory}

\label{firor} In order to set up a perturbative treatment of the master
equation (\ref{me}), it is convenient to decompose the flip rates 
(\ref{eqrates}) into an equilibrium-like contribution 
and a non-equilibrium perturbation 
\begin{eqnarray}
w_{i}(\sigma _{i}\rightarrow -\sigma _{i}) & = & \left[ 1-\frac{\gamma
_{e}+\gamma _{o}}{4}\;\sigma _{i}(\sigma _{i-1}+\sigma _{i+1})\right] 
+\frac{\gamma _{o}-\gamma _{e}}{2} \left[ \frac{(-1)^{i}}{2}\,\sigma
_{i}\,(\sigma _{i-1}+\sigma _{i+1})\right]  \nonumber \\
& \equiv & \bar{w}(\sigma _{i-1},\,\sigma _{i},\,\sigma _{i+1})+
\frac{\gamma_{o}-\gamma _{e}}{2}\Delta _{i}(\sigma _{i-1},\,\sigma _{i},\,\sigma _{i+1}).
\label{split}
\end{eqnarray}
The term in the first $\left[ ...\right] $ bracket, $\bar{w}$, is just the
homogeneous Glauber rate, at an effective temperature defined via $\tanh (2%
\bar{\beta })\equiv \bar{\gamma} \equiv \left( \gamma _{e}+\gamma _{o}\right) /2$. The second
term is explicitly proportional to the ``temperature'' \emph{difference}, 
\begin{equation}
d\equiv \frac{\gamma _{o}-\gamma _{e}}{2}  \label{defexpan}
\end{equation}
and hence captures the non-equilibrium aspect of our dynamics. Thus, $d$
serves as a suitable expansion parameter for a perturbation theory near
equilibrium. We note that it is restricted to the
interval $0\leq \left| d\right| \leq $ $0.5$ where $\left| d\right| =$ $0.5$
corresponds to one temperature being infinite and the other zero. 

We seek the stationary solution $q\left( \{\sigma \}\right) $ of 
Eqn.~(\ref{me}), namely 
\begin{eqnarray}
0=\sum_{j=1}^{N} &&\Bigl[-\left( \bar{w}(\sigma _{j-1},\,\sigma
_{j},\,\sigma _{j+1})+d\;\Delta _{j}(\sigma _{j-1},\,\sigma _{j},\,\sigma
_{j+1})\right) \;q\left( \{\sigma \}\right)  \nonumber \\
&&+\left( \bar{w}(\sigma _{j-1},\,-\sigma _{j},\,\sigma
_{j+1})+d\;\Delta _{j}(\sigma _{j-1},\,-\sigma _{j},\,\sigma _{j+1})\right)
\;q\left( \{\sigma ^{\lbrack j]}\}\right) \Bigr]\quad .  \label{eqstati}
\end{eqnarray}
We assert \cite{js1} that $q\left( \{\sigma \}\right) $ can be written as a
perturbation series, 
\begin{equation}
q\left( \{\sigma \}\right) =\tilde{Z}^{-1} \; q_{0}\left( \{\sigma \}\right) 
\; \leb 1 + \sum_{n=1}^{\infty}d^{n}Q_{n}\left( \{\sigma \}\right) \rib 
\;,  \label{ansatz}
\end{equation} 
where $q_{0}\left( \{\sigma \}\right) \equiv \exp ( \bar{\beta } \; \smsum_i \sig_i
\, \sig_{i+1} )$ is the 
Ising potential at inverse temperature $\bar{\beta }$. Inserting this
ansatz into Eqn.~(\ref{eqstati}), and ordering terms in powers of $d$,
uniqueness demands that the coefficient accompanying each power $d^{n},\; 
n=1,2,...$ vanishes. The lowest order ($d^{0}$) is satisfied by
construction. Higher orders are now evaluated recursively: the coefficient
of $d^{n}$, for $n\geq 1$, can be written as the sum of two terms which must
cancel: $0=S_{1}^{(n)}+S_{2}^{(n)}$. Here, $S_{1}^{(n)}$ contains contributions of
the form $\Delta _{j}Q_{n-1}$ and is therefore explicitly calculable with
the help of results from the ($n-1$)-th order. The second term is generated
by applying the equilibrium rates $\bar{w}(\dots )$ on the unknown $%
Q_{n}$. Even though we cannot invert the Glauber Liouvillean $\bar{w}$
exactly, we can infer the structure of $Q_{n}$, based on the knowledge
of $S_{1}^{(n)}$ and the simplicity of $\bar{w}$ (which contains only
nearest-neighbour spin operators). Of course, at higher orders, the structure
of $S_{1}^{(n)}$ becomes increasingly complex, so that explicit computations
become quite cumbersome. However, at lower orders, this program is quite
feasible. In the following, we illustrate the process for the first-order
correction, and provide a few pointers for the second order.

At order $d$, we need to find $Q_{1}\left( \{\sigma \}\right) $ such that 
\begin{eqnarray}
 0 &=&\sum_{j=1}^{N}\left[ -\Delta _{j}(\sigma _{j-1},\sigma _{j},\sigma
_{j+1})q_{0}\left( \{\sigma \}\right) +\Delta _{j}(\sigma _{j-1},-\sigma
_{j},\sigma _{j+1})q_{0}\left( \{\sigma ^{\lbrack j]}\}\right) \right] 
\nonumber \\
 &+&\sum_{j=1}^{N}\left[ -\bar{w}(\sigma _{j-1},\sigma _{j},\sigma
_{j+1})q_{0}\left( \{\sigma \}\right) Q_{1}\left( \{\sigma \}\right)
+ \bar{w}(\sigma _{j-1},-\sigma _{j},\sigma _{j+1})q_{0}\left(
\{\sigma ^{\lbrack j]}\}\right) Q_{1}\left( \{\sigma ^{\lbrack j]}\}\right) %
\right]  \nonumber \\
&=& S_{1}^{(1)}+S_{2}^{(1)}\; ,  \label{eqpert}
\end{eqnarray}
where $S_{1}^{(1)}$ and $S_{2}^{(1)}$ represent the first and second sum
over $j$, respectively. We first compute $S_{1}^{(1)}$: 
\begin{equation}
S_{1}^{(1)}=-q_{0}\left( \{\sigma \}\right) \frac{\sinh (4\,\bar{\beta }%
)}{2}\sum_{j=1}^{N}\,(-1)^{j}\sigma _{j}\sigma _{j+2}\quad ,  \label{resut1}
\end{equation}
which suggests the following ansatz for $Q_{1}(\{\sigma \})$ 
\begin{equation}
Q_{1}\left( \{\sigma \}\right) =\lambda \sum_{i=1}^{N}(-1)^{i}\sigma
_{i}\sigma _{i+2}\quad ,  \label{ansq1}
\end{equation}
where the (real) parameter $\lambda $ needs to be determined. Inserting (\ref%
{ansq1}) into $S_{2}^{(1)}$, we obtain 
\[
S_{2}^{(1)}=4\lambda \,q_{0}\left( \{\sigma \}\right)
\sum_{j=1}^{N}\,(-1)^{j}\,\sigma _{j}\,\sigma _{j+2}\quad , 
\]
whence 
\begin{equation}
\lambda =-\frac{\sinh (4\,\bar{\beta })}{8}\ .  \label{resula}
\end{equation}
Thus, we arrive at the first order correction to the stationary
distribution: 
\begin{equation}
q\left( \{\sigma \}\right) =\frac{1}{\tilde{Z}}\;q_{0}\left( \{\sigma \}\right)
\left( 1-d\;\frac{\sinh (4\,\bar{\beta })}{8}\sum_{i=1}^{N}(-1)^{i}\,%
\sigma _{i}\,\sigma _{i+2}\right) +{\mathcal{O}}(d^{2})\ .  \label{resuper}
\end{equation}
Hence, in first order perturbation theory the temperature difference induces
an interaction between \emph{next-nearest} neighbours with a sign
difference for even/odd pairs. The correction has a simple, intuitive
interpretation in the defect picture. If $n_{e}(\{\sigma \})$ and $%
n_{o}(\{\sigma \})$ denote the number of \emph{defect pairs} centred on even
or odd sites, respectively, it is easy to see that 
\[
\frac{1}{4}\;\sum_{i=1}^{N}\,(-1)^{i}\,\sigma _{i}\,\sigma
_{i+2}=n_{o}(\{\sigma \})-n_{e}(\{\sigma \})\;, 
\]
i.~e.,~$Q_{1}\left( \{\sigma \}\right) $ is proportional to the \emph{%
difference} between the number of ``even'' and ``odd'' defect pairs. For $%
T_{e}>T_{o}$, $d$ is positive, so that, e.g., a spin configuration with a
single \emph{even} defect pair acquires a \emph{higher} statistical weight
than a configuration with a single \emph{odd} pair. In contrast, those two
configurations are degenerate in equilibrium, i.e, they have the same
weight. Clearly, the high degeneracy of the equilibrium probabilities is
reduced by the perturbative correction: Now, not only the total number of
defects matters, but also how defect pairs are distributed over the two
sub-lattices.

Turning to the \emph{second order} contribution, the calculations, while
becoming more involved, proceed in essentially the same manner. Collecting
all terms of Eq.~(\ref{eqstati}) with a factor $d^2$ the condition for
the correction $Q_2 \leb \{ \sigma \} \rib$ reads 
\bea
0 & =  & \sum_{j=1}^N  \left[ - \Delta_j(\sig_{j-1}, \, 
\sigma_j, \, \sig_{j+1}) \; q_0 \leb \{ \sig \} \rib \; Q_1 \leb \{ \sig \}
\rib + \; \Delta_j(\sig_{j-1}, \, - \sigma_j, \, \sig_{j+1}) 
\; \right. \nonumber  \\
 & &  \times \left. q_0 \leb \{ \sig^{[j]} \} \rib \; Q_1 \leb \{ \sig^{[j]} \} \rib
\; \right] +  \sum_{j=1}^N    \left[  -  \overline{w}(\sig_{j-1}, \, 
\sigma_j, \, \sig_{j+1})) \; q_0 \leb \{ \sig \} \rib 
\; Q_2 \leb \{ \sigma \} \rib \right. \nonumber \\
 & & + \left. \overline{w}(\sig_{j-1}, \, - \sigma_j, \, \sig_{j+1})
 \;  q_0 \leb \{ \sig^{[j]} \} \rib  \;
Q_2 \leb \{ \sigma^{[j]} \} \rib \right] \; =: \; {S}^{(2)}_1 + {S}^{(2)}_2 \; ,
\label{eqpert2}
\eea
where again the two terms ${S}^{(2)}_1 $ and  $  {S}^{(2)}_2 $ 
are defined as the first and the second sum over $j$, respectively.
We obtain
\begin{eqnarray}
&& S_{1}^{(2)} =4\lambda ^{2}q_{0}\left( \{\sigma \}\right) \left\{ \left(
\sum_{i=1}^{N}(-1)^{i}\sigma _{i}\sigma _{i+2}\right) ^{2}\right. 
\nonumber \\
&&+4 \left. \sum_{i=1}^{N}\left[ \left( \sigma _{i}\sigma _{i+2}+\sigma
_{i}\sigma _{i+1}\,\sigma _{i+2}\sigma _{i+3}\right) -\coth (4\,%
\bar{\beta })\left( \sigma _{i}\sigma _{i+1}+\sigma _{i}\sigma
_{i+3}\right) \right] \right\}  \label{t1i2}
\end{eqnarray}
with $\lambda $ given in Eqn.~(\ref{resula}). Again, a suitable ansatz,
based on the structure of $S_{1}^{(2)}$, allows us to determine the
second-order correction $Q_{2}\left( \{\sigma \}\right) $
\begin{eqnarray}
&& Q_{2}\left( \{\sigma  \}\right) =\frac{\lambda ^{2}}{2}\left\{ \left(
\sum_{i=1}^{N}(-1)^{i}\sigma _{i}\,\sigma _{i+2}\right) ^{2} \right.
\nonumber \\
&&+4\left. \sum_{i=1}^{N}\left[ \left( \sigma _{i}\sigma _{i+2}+\sigma
_{i}\sigma _{i+1}\sigma _{i+2}\sigma _{i+3}\right) -\coth (2%
\bar{\beta })\left( \sigma _{i}\sigma _{i+1}+\sigma _{i}\sigma
_{i+3}\right) \right] \right\}  \label{req2} \ .
\end{eqnarray}
Remarkably, we now encounter both, \emph{higher-order} spin operators and 
\emph{next-next-nearest} neighbour interactions.

Clearly, $d^{2}Q_{2}\left( \{\sigma \}\right) $ should be added to the
first order solution, Eqn (\ref{resuper}). It is instructive to
write the stationary distribution in exponential form, via $q\left( \{\sigma
\}\right) \equiv \tilde{Z}^{-1}\exp \left[ \sum_{n=0}^{\infty }d^{n}V_{n}\left(
\{\sigma \}\right) \right] $, to illustrate the differences from the
equilibrium form even more succinctly. To ${\mathcal{O}}(d^{2})$, we obtain: 
\begin{eqnarray}
&&V_{0}\left( \{\sigma \}\right) \equiv \bar{\beta }\sum_{i}\sigma
_{i}\sigma _{i+1}\,,\quad \quad V_{1}\left( \{\sigma \}\right) \equiv
\lambda \sum_{i}(-1)^{i}\,\sigma _{i}\sigma _{i+2}\,, \label{eqpote2}  \\
&&V_{2}\left( \{\sigma \}\right) \equiv 2\lambda ^{2}\sum_{i}\left[ \sigma
_{i}\sigma _{i+2}+\sigma _{i}\sigma _{i+1}\sigma _{i+2}\sigma
_{i+3}-\coth (2\bar{\beta })\left( \sigma _{i}\sigma
_{i+1}+\sigma _{i}\sigma _{i+3}\right) \right]  \nonumber
\end{eqnarray}
The functions $V_{k}$,$\;k=0,\,1,\,...$ contain the interaction terms that
arise in $k$th order of perturbation theory. The system size $N$ enters only
through the summation over $i$. Clearly, the functions $V_{k}$ respect the
basic symmetries of our model: the usual Ising $\mathbb{Z}_{2}$ invariance,
and oddness in $d$ under the exchange of even and odd sub-lattices. Of
course, $V_{0}$ is the Ising Hamiltonian. In stark contrast to the
equilibrium case, the stationary distribution $q\left( \{\sigma \}\right) $
cannot be written as $\exp \left( -\beta H(\{\sigma \})\right) $ with an
appropriate Hamiltonian $H(\{\sigma \})$, since the potentials $V_{k}$
possess a complicated dependence on the two inverse temperatures $\beta _{e}$
and $\beta _{o}$. We also note that, at least for $k\leq 2$, the potentials $%
V_{k}$ are sums of spin operators each of which involves only a local
``cluster'' $S_{k}\equiv \{\sigma _{i},\,\sigma _{i+1},\dots \sigma
_{i+k+1}\}$ of (at most) $k+2$ spins. Hence, it appears that the size of
these clusters grows linearly with $k$, but the coupling strength between
two spins, $\sigma _{i}$ and $\sigma _{i+k+1}$, decreases exponentially as $%
d^{k}$. We conjecture that this behaviour lies at the source of the
exponential decay which characterises correlation functions in this model %
\cite{raczzia,bsfs}.

We conclude this section with some comments on the accuracy of our
perturbation theory. The effective expansion parameter, up to and including
second order, is $\lambda d$, so that the mean inverse temperature, $%
\bar{\beta }$, also plays a role. In particular, the limit $\bar{%
\beta }\rightarrow \infty $ looks troubling at first sight, since $\lambda
\propto \sinh (4\,\bar{\beta })$ (cf. Eqn.~(\ref{resula})). However, a
more careful analysis shows that $\left| \lambda d\right| <1$ independently
of how $\gamma _{o}$ and $\gamma _{e}$ approach unity, and our perturbation
theory remains valid for low mean temperature. 
\begin{figure}[tbp]
\centerline{\epsfig{figure=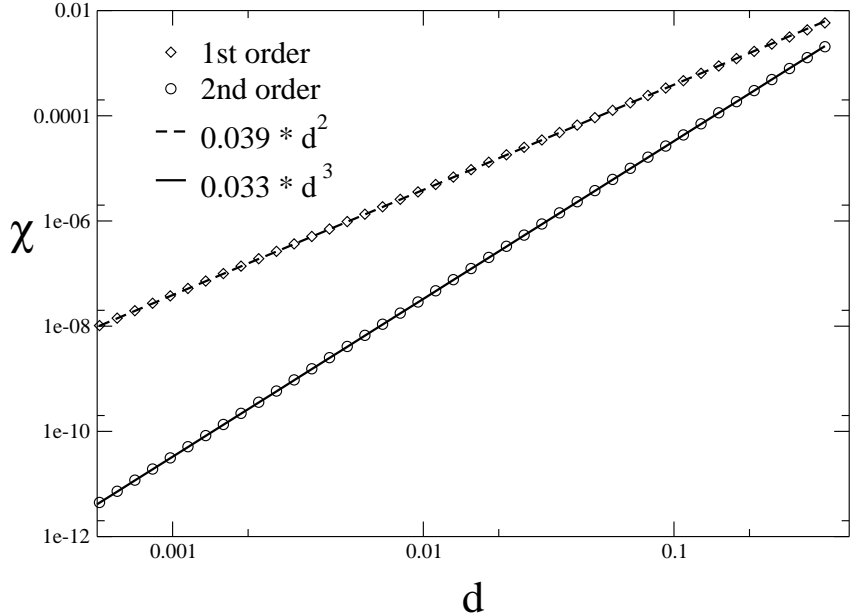, width=100mm, angle=-90 }}
\caption{Quantity $\protect\chi$, from Eqn.~(\ref{defchi}), 
evaluated for a chain of 6 sites for first (diamonds) and second (circles) order 
perturbation theory. We plot $\chi$ as a function of $d$, at fixed mean
temperature $\bar{\gamma}=0.4$.}  
\label{figchi}
\end{figure}

To test its numerical accuracy, we compare our perturbative solution, $%
q_{pert}(\{\sigma \})$, with an exact result, $q_{ex}(\{\sigma \})$, for a
short chain of $6$ sites \cite{Irina}. For such small systems, the zero
eigenvector (stationary distribution) of the master equation is easily
found. As a quantitative measure, we define 
\begin{equation}
\chi \equiv \sqrt{\sum_{\{\sigma \}}\left( q_{pert}(\{\sigma
\})-q_{ex}(\{\sigma \})\right) ^{2}}\quad .  \label{defchi}
\end{equation}%
Keeping the mean temperature, i.e. $(\gamma _{e}+\gamma _{o})/2$, fixed, we
expect $\chi $ to scale as $\chi \sim d^{k+1}$ for $d\rightarrow 0$, if we
consider $q_{pert.}$ to $k$th order. Figure \ref{figchi} confirms this
scaling behaviour for $k=1$ and $k=2$, at\ $(\gamma _{e}+\gamma _{o})/2=0.4$%
, for a large range of $d$. Remarkably, even for the largest $d$
investigated ($0.4$), $\chi $ remains well below $0.01$, indicating that our
perturbative solution is an excellent approximation to the exact result. For
smaller $d$, the deviations are even smaller, giving us considerable
confidence in our perturbative approach.

In the next two sections, we turn to two general features 
of our two-temperature spin
chain, namely the violation of detailed balance and the concept of entropy
production.

\section{Violation of detailed balance}

\label{secdetb} Following \cite{muka}, it is straightforward to show that
our dynamics violates detailed balance, even without knowing
the steady-state distribution. The proof involves closed orbits (loops) in
configuration space, $\{\sigma \} _{1} \rightarrow \{\sigma \} _{2} \rightarrow
\dots \rightarrow \{\sigma \} _{k} \rightarrow \{\sigma \} _{1}$, of ``length'' $%
k=1,2,..$. Each of the configurations in the loop differs from its successor
by a single spin flip. For each closed orbit we consider the product of the
transition rates, traversing the loop in the ``forward'' and ``reverse''
direction, namely 
\begin{eqnarray*}
\Pi _{f} &\equiv &w\left( \{\sigma \} _{k} \rightarrow \{\sigma \} _{1} \right)
\prod_{j=1}^{k-1}w\left( \{\sigma \} _{j} \rightarrow \{\sigma \} _{j+1} \right)
\; , \quad \\
\Pi _{r} &\equiv &w\left( \{\sigma \} _{1} \rightarrow \{\sigma \} _{k} \right)
\prod_{j=1}^{k-1}w\left( \{\sigma \} _{j+1} \rightarrow \{\sigma \} _{j}
\right) \; ,
\end{eqnarray*}
respectively. Detailed balance is 
violated, if we can identify a single loop for
which $\Pi _{f}\neq \Pi _{r}$. For our model, the shortest loop of this type
has length $4$ and involves four nearest-neighbour spins, $\sigma
_{i-1},...\sigma _{i+2}$. We begin with a configuration $\{\sigma \} _{1}$ 
in which these four
spins are all $+1$, and $i$ is even. The two central spins play the key
role: $\{\sigma \} _{2}$ has spin $\sigma _{i}=-1$; $\{\sigma \} _{3}$ has $%
\sigma _{i}=-1$ and $\sigma _{i+1}=-1$; finally, $\{\sigma \} _{4}$ has $%
\sigma _{i+1}=-1$. Flipping $\sigma _{i+1}$ back to $+1$ restores $\{\sigma
\} _{1}$. In defect language, we first create an even defect pair, then the
right defect, followed by the left one, diffuses right by one lattice site,
and then the resulting odd defect pair is annihilated: $000\rightarrow
110\rightarrow 101\rightarrow 011\rightarrow 000$. It is easy to compute $%
\Pi _{f}=(1-\gamma _{e})(1+\gamma _{o})$ which differs from $\Pi
_{r}=(1+\gamma _{e})(1-\gamma _{o})$. Thus our model does indeed violate
detailed balance.

Our (perturbative) steady state solution sheds further light on this issue,
demonstrating which, and how, elementary spin flips contribute. We focus on
the net probability current $F$ between two configurations, $\{\sigma \}$ and $%
\{\sigma ^{\lbrack i]}\}$, 
\begin{equation}
F\left( \{\sigma \},\{\sigma ^{\lbrack i]}\}\right) \equiv w_{i}(\sigma
_{i}\rightarrow -\sigma _{i})q(\{\sigma \})-w_{i}(-\sigma
_{i}\rightarrow \sigma _{i})q(\{\sigma ^{\lbrack i]}\})\;.  \label{flux}
\end{equation}
In \emph{zeroth} order of perturbation theory, all $F$ 
vanish since detailed balance holds.
To obtain the leading \emph{non-vanishing} contributions to $F$, we
decompose the rates according to Eqn. (\ref{split}) and insert the first
order expression for the stationary density. To ${\mathcal{O}}(d)$, we find 
\[
\hspace{-1cm}F\left( \{\sigma \},\{\sigma ^{\lbrack i]}\}\right)
=\tilde{Z}^{-1}d\left( A_{1}+A_{2}\right) q_{0}\left( \{\sigma \}\right) +
{\mathcal{O}}(d^{2}) 
\]
with 
\begin{eqnarray*}
A_{1} &\equiv &\Delta _{i}(\sigma _{i-1},\sigma _{i},\sigma
_{i+1})-\Delta _{i}(\sigma _{i-1},-\sigma _{i},\,\sigma _{i+1})\frac{%
q_{0}\left( \{\sigma ^{\lbrack i]}\}\right) }{q_{0}\left( \{\sigma \}\right) 
} \; , \\
A_{2} &\equiv &\bar{w}(\sigma _{i-1},\sigma _{i},\sigma
_{i+1})Q_{1}\left( \{\sigma \}\right) -\bar{w}(\sigma _{i-1},-\sigma
_{i},\sigma _{i+1})\frac{q_{0}\left( \{\sigma ^{\lbrack i]}\}\right) }{%
q_{0}\left( \{\sigma \}\right) }Q_{1}(\{\sigma ^{\lbrack i]}\}) \quad .
\end{eqnarray*}
Here, $A_{1}$ reflects the effect of the non-equilibrium rate, $\Delta$, whereas 
$A_{2}$ contains the first-order correction, $Q_{1}\left( \{\sigma \}\right) $%
, to the Ising stationary distribution. Using our definitions and results
from the previous section, one finds easily that 
\begin{equation}
A_{1}=\frac{(-1)^{i}}{2}\sigma _{i}\,(\sigma _{i-1}+\sigma _{i+1})\left[
1+\exp \left( -2\bar{\beta }\sigma _{i}\,(\sigma _{i-1}+\sigma
_{i+1})\right) \right] \quad  \label{resua1}
\end{equation}
and 
\begin{eqnarray}
A_{2} &=&\frac{\sinh (4\bar{\beta })}{2}\left[ 1-\frac{\bar{%
\gamma }}{2}\sigma _{i}\ (\sigma _{i-1}+\sigma _{i+1})\right] \nonumber \\ 
& & \times  \left\{ \left[ n_{o}(\{\sigma ^{\lbrack i]}\})-n_{e}(\{\sigma ^{\lbrack
i]}\})\right] -\left[ n_{o}(\{\sigma \})-n_{e}(\{\sigma \})\right] \right\} .
\label{resua2}
\end{eqnarray}
One can show, by considering elementary spin flips and their effect on the
defect representation, that the $\left\{ ...\right\} $ bracket in Eqn. (\ref%
{resua2}) can take only the values $\pm 1$ or $0$. 

We are now ready to evaluate the relevance of different spin flips types (cf. the
discussion following Eqn.~(\ref{me})) for detailed balance violation.
Spin flips of type (a) and (b) are inverse to one another: where (b)
generates a defect pair, (a) annihilates it. To be specific, we let $%
\{\sigma \}\rightarrow \{\sigma ^{\lbrack i]}\}$ be a type (b) transition,
with $\sigma _{i-1}=\sigma _{i}=\sigma _{i+1}$ in $\{\sigma \}$. For this
configuration, we find 
\begin{eqnarray}
A_{1} &=&(-1)^{i}\left[ 1+\exp \left( -4\bar{\beta }\right) \right]
\quad ,  \nonumber \\
A_{2} &=&\frac{1}{2}\left\{ \left[ n_{o}(\{\sigma ^{\lbrack
i]}\})-n_{e}(\{\sigma ^{\lbrack i]}\})\right] -\left[ n_{o}(\{\sigma
\})-n_{e}(\{\sigma \})\right] \right\} \quad .  \label{eqmaga1}
\end{eqnarray}
Given the constraint on the $\left\{ ...\right\} $ bracket, the inequality $%
\left| A_{2}\right| <2\left| A_{1}\right| $ follows. Thus, $A_{1}$
determines the \emph{sign} of the probability current. For $%
T_{e}>T_{o}\;(d>0)$ we have $F\left( \{\sigma \},\,\{\sigma ^{\lbrack
i]}\}\right) >0$ if $i$ is even, and negative otherwise: At the hotter
(even) sites, defect pairs are more often generated than removed. At the
cooler (odd) sites, the situation is reversed. In this manner, spin flips of
types (a) and (b) always contribute to detailed balance violation in the
presence of two temperatures.

If the spin flip $\{\sigma \}\rightarrow \{\sigma ^{\lbrack i]}\}$ is a type
(c) process (i.e., defect diffusion\thinspace ), then its inverse also
belongs to this class, and both occur with rate $1$. Since $%
\sigma_{i-1}=-\sigma _{i+1}$ for a type (c) spin flip, $A_{1}$ vanishes and
only $A_{2}$ contributes to the probability current. As a result, a type (c)
transition violates detailed balance in first 
order of $d$ (i.e., generates a nonzero $F$) only if
it changes the difference of odd and even defect pairs. In defect notation,
the process $...0,1,1,0,0,...$ $\rightarrow $ $...0,1,0,1,0,...$violates
detailed balance while $...0,1,0,0,...$ $\rightarrow $ $...0,0,1,0,...$ does
not. If the defect pair in the former process is centred on an even site, we
will have $F\left( \{\sigma \},\,\{\sigma ^{\lbrack i]}\}\right) >0$ for $%
T_{e}>T_{o}$, etc.

Including terms of ${\mathcal{O}}(d^{2})$, we find that \emph{all} type (c) spin flips now
generate non-vanishing probability currents. Thus, 
currents between pairs of configurations, connected by single spin flips,
are generically nonzero.

\section{Entropy production}

\label{secentr} We finally consider two different approaches to entropy
production in our model. The first \cite{raczzia} is very intuitive and
tailored to our specific model, by focusing on the energy (``heat'') flux
from one bath to the other. The second method \cite%
{maes1,lebspo,maes2,schnake} is more general, defining an entropy production
for any master equation. In the following, we show that both approaches are
equivalent for our model.

We briefly review the first method. As the spin chain is connected to two
baths at different temperatures, heat flows from the warmer reservoir into
the spin chain and on into the colder one. Since the energy difference of
two configurations connected via a single spin flip at site $i$ is $%
-2\,\sigma _{i}\,(\sigma _{i-1}+\sigma _{i+1})$, the energy transfer at this
site is given by 
\begin{equation}
j(i)\equiv \langle -2\,\sigma _{i}\,(\sigma _{i-1}+\sigma
_{i+1})\;w_{i}(\sigma _{i}\rightarrow -\sigma _{i})\;\rangle  \label{heattr}
\end{equation}
where $\left\langle \cdot \right\rangle $ denotes the configurational
average with respect to $q(\{\sigma \})$. Standard thermodynamics then
suggests to define the entropy production of the whole chain as 
\begin{equation}
\dot{S}=\frac{N}{2}\;\left( \frac{j\,(2\,i+1)}{T_{o}}+\frac{j\,(2\,i)}{T_{e}}%
\right) \quad .  \label{entterm}
\end{equation}
Using Eqn.~(\ref{eqrates}) for the rates, $j(i)$ is easily expressed in
terms of the \emph{exactly known} \cite{raczzia} two spin correlation
function 
\bea
\langle \sigma _{i} \,\sigma _{j} \rangle & = &  
\sqrt{ A(i) \ A(j) } \; \;  \lambda^{\left| j-i\right| } \quad \nonumber  \\ 
{\rm with} \quad A(i) & \equiv & \; \left\{ 
\begin{array}{l}
\;A _{e}=(\gamma_e + \gamma_o ) / (2 \; \gamma_o) \,,\quad \quad i\;\;{even} 
\\ 
\;A _{o}=(\gamma_e + \gamma_o ) / (2 \; \gamma_e) \,,\quad \quad i\;\;{odd}
\end{array} \right. \; , \nonumber \\
\hspace{1.7cm} \lambda & \equiv & \frac{1}{\sqrt{\gamma_e \;  \gamma_o}} \;
\leb 1 - \sqrt{1 - \gamma_e \;  \gamma_o } \rib \quad .
\eea
Hence,
\begin{equation}
\dot{S}=\frac{N}{2}\left( \gamma _{o}-\gamma _{e}\right) \left( \frac{1}{%
T_{o}}-\frac{1}{T_{e}}\right)  \label{entrz}
\end{equation}
follows exactly, for any choice of temperatures. We note that the entropy
production is positive whenever the two temperatures differ.

The more general approach \cite{maes1,lebspo,maes2,schnake} to entropy
production begins with the Gibbs entropy for the full time-dependent
probability distribution $p\left( \{\sigma \},\,t\right) $: 
\begin{equation}
S_{G}(t)=-\sum_{\{\sigma \}}\,p\left( \{\sigma \},\,t\right) \;\ln p\left(
\{\sigma \},\,t\right) \quad .
\end{equation}
For the stationary distribution $q\left( \{\sigma \}\right) $, the time
derivative $dS_{G}/dt$ must vanish. With the help of the master equation, we
find $dS_{G}/dt=\dot{\tilde{S_{1}}}-\dot{\tilde{S_{2}}}$ where 
\begin{eqnarray}
\dot{\tilde{S_{1}}} &=&   \frac{1}{2} \; \sum_{\{\sigma \}}\sum_{i=1}^{N}\left[ w_{i}(\sigma
_{i}\rightarrow -\sigma _{i})\;q(\{\sigma \})-w_{i}(-\sigma _{i}\rightarrow
\sigma _{i})\;q(\{\sigma ^{\lbrack i]}\})\right]  \nonumber \\
&&\times \ln \left( \frac{w_{i}(\sigma _{i}\rightarrow -\sigma
_{i})\;q(\{\sigma \})}{w_{i}(-\sigma _{i}\rightarrow \sigma
_{i})\;q(\{\sigma ^{\lbrack i]}\})}\right) \; ,  \nonumber \\
\dot{\tilde{S_{2}}} &=&\sum_{\{\sigma \}}\sum_{i=1}^{N}q(\{\sigma
\})w_{i}(\sigma _{i}\rightarrow -\sigma _{i})\ln \left( \frac{w_{i}(\sigma
_{i}\rightarrow -\sigma _{i})}{w_{i}(-\sigma _{i}\rightarrow \sigma _{i})}%
\right) \quad .
\label{formaes}
\end{eqnarray}
The two expressions $\dot{\tilde{S_{1}}}$ and $\dot{\tilde{S_{2}}}$
constitute two alternate representations for the entropy production. In
particular, we recognise the probability current $F(\{\sigma \},\,\{\sigma
^{\lbrack i]}\})$ of Eq.~(\ref{flux}) in 
the expression for $\dot{\tilde{S_{1}}}$. One easily
sees $\dot{\tilde{S_{1}}}\geq 0$ since each term in the sum is non-negative.
Moreover, $\dot{\tilde{S_{1}}}$ vanishes if and only if detailed balance is
satisfied. Thus, strictly positive entropy production is equivalent to the
violation of detailed balance.

We now evaluate the expression for $\dot{\tilde{S_{2}}}$ in Eqn.~(\ref%
{formaes}) \cite{js2}. Since our model exhibits
translation invariance with period $2$, we only have to consider one even ($%
2\,i$) and one odd ($2\,i+1$) site, so that 
\begin{equation}
\dot{\tilde{S_{2}}}=\frac{N}{2}\sum_{\{\sigma \}}\;\sum_{j\in
\{2\,i,\,2\,i+1\}}q(\{\sigma \})w_{j}(\sigma _{j}\rightarrow -\sigma _{j})%
\left[ -2\beta _{j}\sigma _{j}(\sigma _{j-1}+\sigma _{j+1})\right] \ .
\end{equation}
Here, we have used the simple relation 
\[
\frac{w_{i}(\sigma _{i}\rightarrow -\sigma _{i})}{w_{i}(-\sigma
_{i}\rightarrow \sigma _{i})}=\exp \left[ -2\beta _{i}\sigma _{i}\left(
\sigma _{i-1}+\sigma _{i+1}\right) \right] 
\]
Recalling Eqn. (\ref{heattr}), we see immediately that 
\begin{equation}
\dot{\tilde{S_{2}}}=\frac{N}{2}\;\left[ \beta _{e}j(2\,i)+\beta _{o}j(2\,i+1)%
\right] \quad 
\end{equation}
which is exactly the expression (\ref{entterm}) that was derived from
thermodynamics. Thus, we confirm the consistency of the more abstract
general expressions (\ref{formaes}) and the intuitive thermodynamic
approach. While this enhances our confidence in (\ref{formaes}), a better
understanding in terms of more basic physical principles is still
outstanding.

\section{Conclusions}

\label{seccon} To summarise, we have gained some analytic insight into the
steady state configurational probabilities of an Ising spin chain driven out
of equilibrium by a coupling to two heat baths: Subject to different
temperatures, spins on even and odd sites are updated according to a
generalisation of the usual Glauber rates. A perturbative calculation in the
temperature difference shows that the stationary distribution for this
non-equilibrium model is rather complex, with longer-range and higher-order
spin operators appearing. Clearly, the energy of a configuration 
no longer determines its statistical weight.  
Unfortunately, at this stage we can only
conjecture the structure of this distribution beyond second order in
perturbation theory. It seems likely that, at each order, additional spin
operators will have to be introduced. Turning to other characteristics of
this non-equilibrium steady state, we could  understand in more detail how
the presence of two temperatures violates detailed balance. We could also
show the equivalence of two unrelated definitions of entropy production,
thus providing additional support for the general expression \cite%
{maes1,lebspo,maes2,schnake}.

Of course, numerous questions remain open. In particular, we would like to
reconcile the apparent proliferation of longer-range spin couplings in the
configurational probabilities with the observed ``triviality'' of
long-wavelength properties. At present, we have strong  indications \cite%
{bsfs} that arbitrary spin correlation functions decay exponentially, quite
similar to the equilibrium Ising chain. However, the most fundamental
question still remains unanswered: are there any a priori criteria, beyond
simple symmetries, that determine which configurations will appear with
equal weights? In equilibrium, the determining quantity is their
energy. Far from equilibrium, the key concept is still missing.  

\textbf{Acknowledgements}
We thank J. Slawny, I. Mazilu, R.K.P. Zia, U.C. T\"{a}uber, W.~Just,
E. Ben-Naim and M. Hastings for fruitful discussions. Financial support 
from the DFG through contract no. SCHM 1537 and from the NSF through 
the Division of Materials Research is gratefully acknowledged.

\end{document}